\documentclass[aps,prd,twocolumn,superscriptaddress,longbibliography,nofootinbib]{revtex4-2}
\usepackage[utf8]{inputenc}
\usepackage{color}
\usepackage{graphicx}
\usepackage{subfigure}
\usepackage{xcolor}
\usepackage[normalem]{ulem}
\usepackage[pdftex,breaklinks,colorlinks,linkcolor=blue,citecolor=teal,anchorcolor=red,urlcolor=cyan]{hyperref}
\usepackage{orcidlink}

\usepackage{amsmath,amssymb,amsfonts}

\def\prl{Phys. Rev. Lett.}
\def\prd{Phys. Rev. D}
\def\cqg{Class. Quantum Grav.}

\def\pau_p{Prog. Theor. Phys.}

\def\prague{{\tt prague}}
\def\sphGR{{\tt sphGR}}
\def\bamps{{\tt bamps}}

\begin{document}

\title{Critical phenomena in the collapse of gravitational waves}

\author{Thomas W.~Baumgarte\orcidlink{0000-0002-6316-602X}}
\affiliation{Department of Physics and Astronomy, Bowdoin College, Brunswick, ME 04011, USA}

\author{Bernd Brügmann\orcidlink{0000-0003-4623-0525}}
\affiliation{Friedrich-Schiller-Universität,
Jena, 07743 Jena, Germany}

\author{Daniela Cors\orcidlink{0000-0002-0520-2600}}
\affiliation{Friedrich-Schiller-Universität,
Jena, 07743 Jena, Germany}

\author{Carsten Gundlach\orcidlink{0000-0001-9585-5375}}
\affiliation{School of Mathematical Sciences, University of Southampton,
  Southampton SO17 1BJ, United Kingdom} 

\author{David~Hilditch\orcidlink{0000-0001-9960-5293}}
\affiliation{
  Centro de Astrof\'{\i}sica e Gravita\c c\~ao -- CENTRA,
  Departamento de F\'{\i}sica, Instituto Superior T\'ecnico -- IST,
  Universidade de Lisboa -- UL, Av.\ Rovisco Pais 1, 1049-001 Lisboa,
  Portugal}
 
\author{Anton Khirnov\orcidlink{0000-0002-9569-381X}}
\affiliation{
Institute of Theoretical Physics, Faculty of Mathematics and Physics, Charles University, CZ-180 00 Prague, Czech Republic
}

\author{Tom\'a\v{s} Ledvinka\orcidlink{0000-0002-6341-2227}}
\affiliation{
Institute of Theoretical Physics, Faculty of Mathematics and Physics, Charles University, CZ-180 00 Prague, Czech Republic
}

\author{Sarah Renkhoff\orcidlink{0000-0002-1233-2593}}
\affiliation{Friedrich-Schiller-Universität,
Jena, 07743 Jena, Germany}

\author{Isabel Su\'arez Fern\'andez\orcidlink{0000-0002-6725-5759}}
\affiliation{
  Departament de Física, Universitat de les Illes Balears, IAC3 – IEEC, Crta. Valldemossa km 7.5, E-07122 Palma, Spain}

\begin{abstract}
Fine-tuning generic but smooth spherically-symmetric initial data for general relativity to the threshold of dynamical black hole formation creates arbitrarily large curvatures, mediated by a universal self-similar solution that acts as an intermediate attractor. For vacuum gravitational waves, however, these critical phenomena have been elusive. We present, for the first time, excellent agreement among three independent numerical simulations of this collapse. Surprisingly, we find no universality, and observe approximate self-similarity for some families of initial data but not for others.
\end{abstract}

\maketitle

%=========================================================
%\section{Introduction}
%\label{sec:intro}
%=========================================================

Critical phenomena in gravitational collapse were first reported in the seminal work of Choptuik \cite{Cho93}. Specifically, Choptuik considered a massless scalar field minimally coupled to general relativity in spherical symmetry.  For different families of initial data, each parametrized by some $p$, he noted that each one can be fine-tuned to a critical parameter $p_*$ that separates subcritical data, i.e.~those that disperse and leave behind flat space, from supercritical data that form black holes.  In the vicinity of $p_*$, Choptuik observed critical phenomena with remarkable similarity to those observed in other fields of physics.  In particular, the mass of black holes formed in the collapse of supercritical data scales with
\begin{equation} \label{mass_scaling}
M \simeq | p - p_* |^{\gamma},
\end{equation}
where the critical exponent $\gamma$ is {\em universal} to the matter model considered, and hence independent of the family of initial data.  Fine-tuning to $p_*$, the dynamical evolution approaches a {\em self-similar} critical solution that is again universal.  

Choptuik's original announcement triggered a large body of work that established critical phenomena in gravitational collapse for different matter models, dimensions, symmetry assumptions, and asymptotics (see \cite{GunM07} for a review), resulting in a thorough understanding of these phenomena at least in the context of spherical symmetry. Depending on the matter model, the critical solution turns out to be either {\em discretely} self-similar (DSS, for example for the scalar field considered by Choptuik) or {\em continuously} self-similar (CSS, for example for radiation fluids \cite{EvaC94}).  The critical exponent $\gamma$ is the inverse of the Lyapunov exponent of the single unstable perturbation mode of the critical solution \cite{KoiHA95,Mai96}, relating the universality of the critical exponent to that of the critical solution.  In spherical symmetry the accumulation point, the spacetime event towards which the self-similar solution contracts, must be located at the center of the symmetry.

Abrahams and Evans (\cite{AbeE93,AbeE94}, hereafter AE) presented evidence for critical phenomena in {\em vacuum} (gravitational-wave) collapse very soon after Choptuik's announcement, specifically in twist-free axisymmetry with an additional reflection symmetry through the equator.  Fine-tuning two families of initial data to the onset of collapse, they reported a scaling exponent $\gamma\simeq 0.36$ and echos in the gravitational field, ``lending support to the contention" that there exists a unique, DSS critical solution \cite{AbeE94}. 

Though a number of authors (e.g.~\cite{AlcABLSST00,GarD01,Rin08a,Sor10,Hiletal13}) have performed simulations of nonlinear gravitational waves using a number of different approaches, it has been difficult to reproduce the results of AE.  Some of these attempts were hampered by numerical problems, others found some evidence for scaling but with different scaling exponents, and none have been able to establish the existence of a universal self-similar critical solution.

In this Letter we suggest that the expectation of a universal critical solution with an exact DSS in the collapse of gravitational waves -- and therefore in the absence of spherical symmetry in general -- is not supported by the currently available numerical results. Rather, different sets of initial data may result in different threshold solutions. Some of these show an approximate DSS in our simulations, but others do not.

We start by briefly summarizing recent numerical evidence for the absence of a universal critical solution in the collapse of gravitational waves, based on independent simulations.  The authors of \cite{LedK21} use two common approaches for constructing gravitational-wave initial data.  One such approach, similar to that adopted by AE, starts with {\em Teukolsky wave} \cite{Teu82} solutions to the linearized Einstein equations, and solves Einstein's constraint equations in order to obtain valid non-linear initial data.  A second, {\em Brill wave}, family instead adopts the procedure of \cite{Bri59}. (See also \cite{SuaBH21} for a comparison of the two approaches at the linear level.)  Both approaches involve choosing a seed function that determines the shape and location of the initial wave package; in particular, this seed function can be positive or negative, and can describe a ``centered" or ``off-center" wave.  \cite{LedK21} found different critical exponents $\gamma$ for these different families. Similar results were found by \cite{HilWB17,SuaRCBH22} for different families of Brill waves.  Finally, \cite{BauGH23} compared quadrupolar and hexadecapolar Teukolsky waves (see the generalization of \cite{Rin08b} for arbitrary multipole moments) and found both quantitative and qualitative differences in the corresponding threshold solutions.   

In generic non-spherical initial data, one expects a single accumulation point of echos (for a given threshold value $p_*$), whose location is not known a priori. For generic axisymmetric data, this must be on the axis. For axisymmetric data with an additional equatorial reflection symmetry, there could be one accumulation point at the center, or two located symmetrically on the axis.  Evidence for two separate ``centers of collapse" was first provided by \cite{ChoHLP03b}, who considered aspherical deformations of scalar fields and observed a ``bifurcation" of the collapse region for large departures from spherical symmetry and exquisite fine-tuning (see also \cite{Bau18}).  Similarly, \cite{HilWB17,LedK21,SuaRCBH22} observed that fine-tuning equatorially symmetric families of Brill wave initial data to the onset of collapse resulted in two separate centers of collapse.  The two centers, one above and one below the center on the symmetry axis, are identified both by extrema of the Kretschmann curvature scalar
\begin{equation} \label{Kretschmann}
    I = R_{abcd} R^{abcd},
\end{equation}
and by the formation of two separate apparent horizons.  Similar findings were reported by \cite{BauGH23} for hexadecapolar Teukolsky waves, but {\em not} for quadrupolar Teukolsky waves, for which the maxima of $I$ occur at the center. (See also \cite{BauGH19} and \cite{PerB21} for similar behavior observed in the gravitational collapse of dipolar versus quadrupolar electromagnetic waves.)

If there are two ``centers", in the sense of the location of recurring curvature maxima, they could arise from DSS with a single accumulation point (with two locations of local curvature maxima in a single critical solution meeting at the accumulation point), but with equatorial reflection symmetry they could also arise from two separate accumulation points located symmetrically, with the same $p_*$ by symmetry.  The former was reported by \cite{BauGH23} for hexadecapolar Teukolsky waves.

Here we employ three different formulations, gauges, and codes to further analyze this situation. \prague~\cite{LedK21} is a finite-difference code based on the Einstein Toolkit \cite{ET12} that solves the Baumgarte-Shapiro-Shibata-Nakamura (BSSN) formulation of Einstein's equations \cite{NakOK87,ShiN95,BauS98} using a ``quasi-maximal" slicing condition (see \cite{KhiL18}).  \bamps~\cite{HilWB16,Bru11} is a pseudo-spectral code that solves Einstein's equations in a first-order generalized harmonic formulation \cite{LinSKOR06} with gauge conditions and refinement strategy as discussed in \cite{HilWB16, RenCH23}.  Finally, \sphGR~\cite{BauGH23} is a finite-difference BSSN code in spherical polar coordinates (see \cite{BauMCM13}), and uses the shock-avoiding slicing condition suggested by \cite{Alc97} (see also \cite{BauH22} for a comparison with 1+log slicing).  While \prague\ and \bamps\ use mesh refinement to resolve increasingly small features, \sphGR\ is less well suited for attaining the necessary resolution away from the center (see \cite{BauGH23} for a discussion).  Accordingly, we do not include results from \sphGR\ for initial data close to the black-hole threshold. 

\begin{figure}
    \centering
    \includegraphics[width = 0.48 \textwidth]{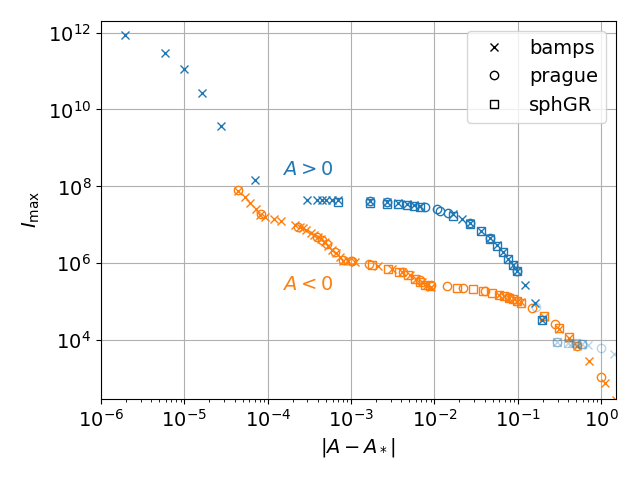}
    \caption{The maximum $I_{\rm max}$ of the Kretschmann scalar (\ref{Kretschmann}) encountered in subcritical evolutions of Brill waves with the seed function (\ref{Brill_seed}), as a function of the amplitude $A$ (see also Fig.~1 in \cite{LedK21} and Fig.~5 in \cite{SuaRCBH22}). The faded data for $A > 0$ with $A_* - A \gtrsim 0.2$ mark maxima that occur in the initial data.}
    \label{fig:Brill_comparison}
\end{figure}

\begin{figure*}
    \centering
    \includegraphics[scale = .83]{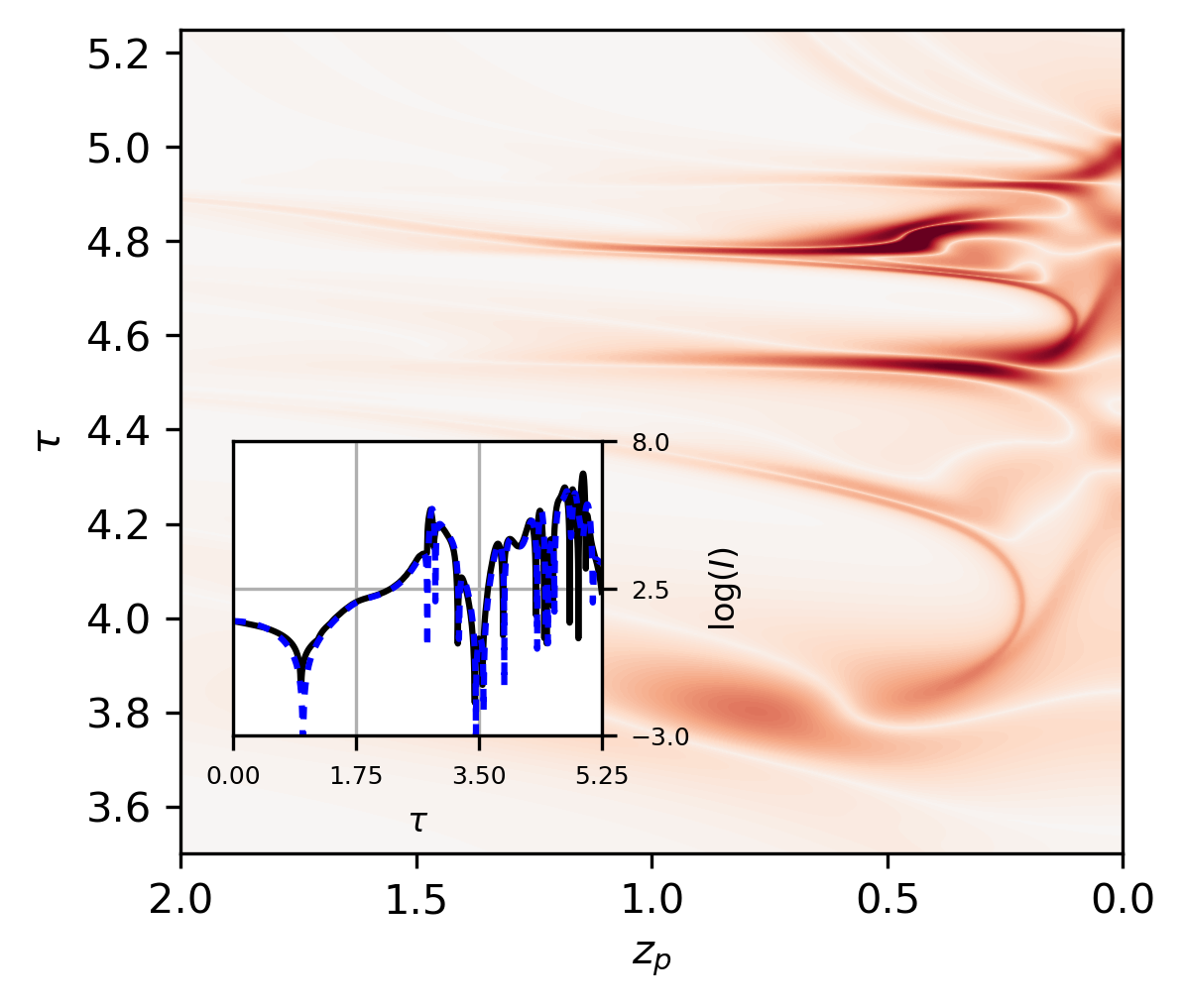} \hspace{-0.2cm}
    \includegraphics[scale = .83]{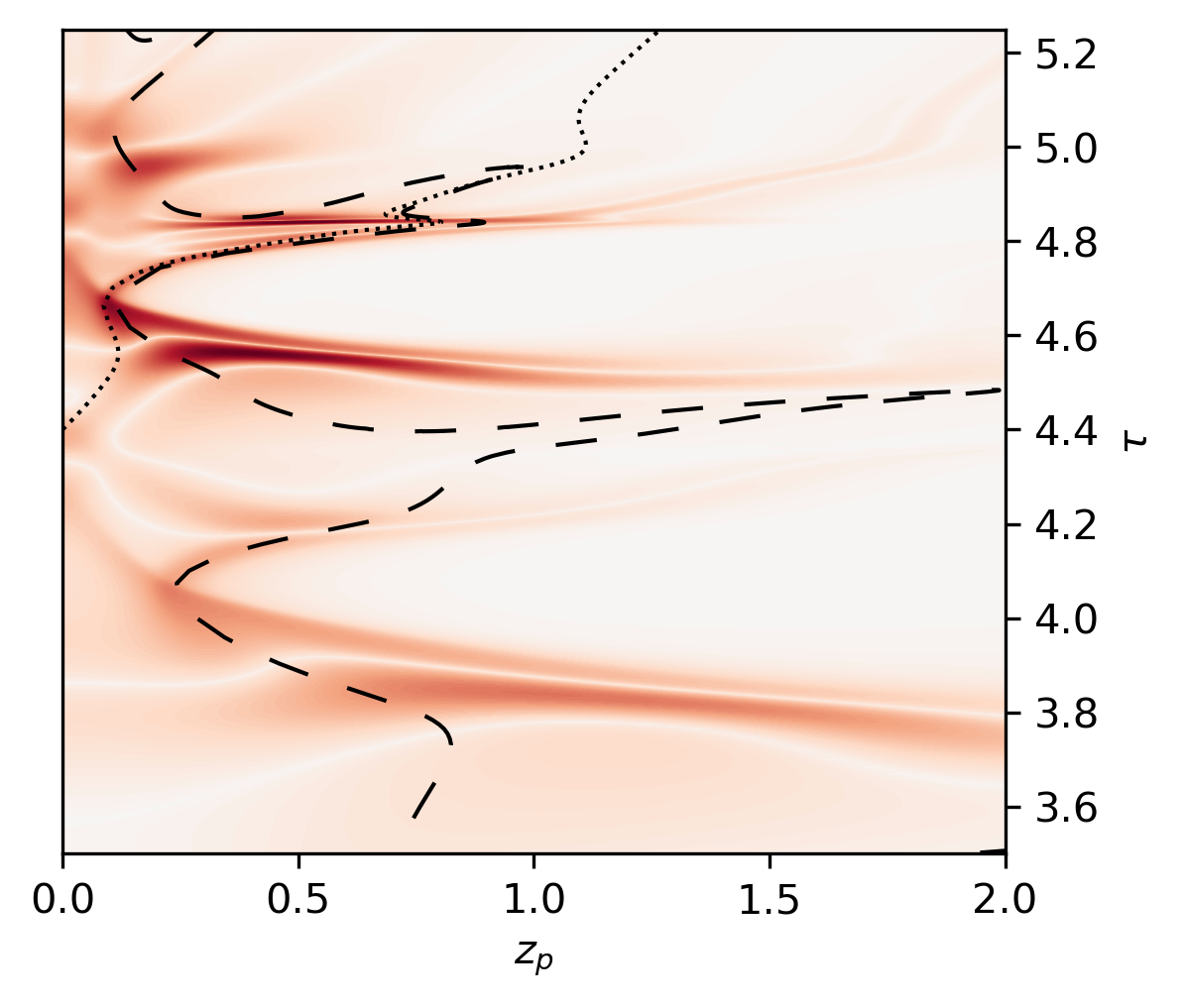}\\
    \includegraphics[scale = .83]{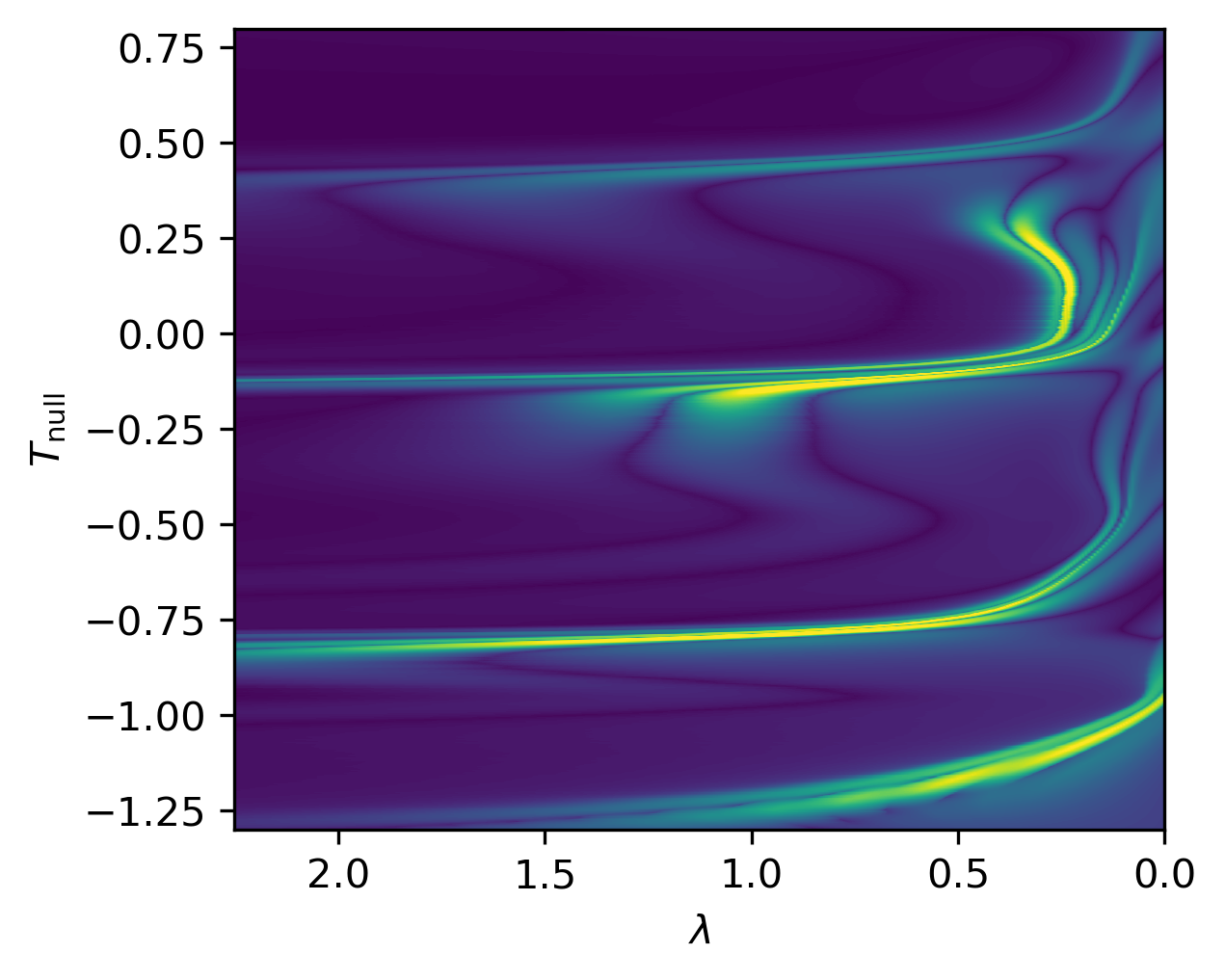}    \hspace{-0.2cm}
    \includegraphics[scale = .83]{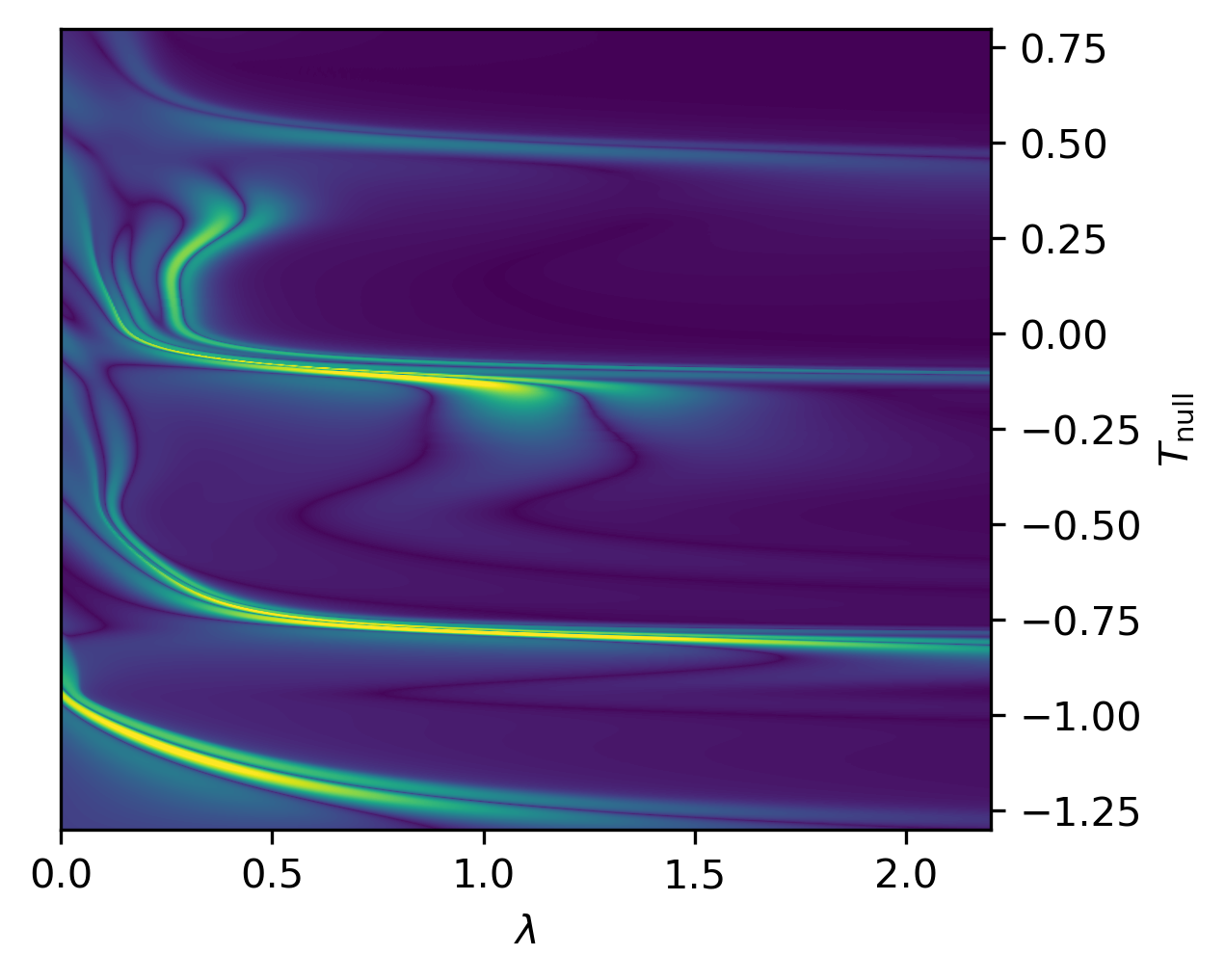}
    
    \caption{Color plots of the Kretschmann scalar $I^{1/4}$ on the symmetry axis, for a near-critical evolution of Brill initial data with the (negative $A$) seed function (\ref{Brill_seed}).  The two panels on the left show \bamps\ results with $A=-3.50909$, and the two on the right \prague\ results with $A=-3.5090625$. The two top panels show results along the time slices of each code, as a function of proper distance $z_p$ along the slice from the center and proper time $\tau$ at the center on that slice. This presentation is still slicing-dependent. The inset in the top left shows $\log_{10} I$ at the center versus proper time $\tau$ for both the \bamps\ (solid black line) and \prague\ (dashed blue line) data. The two bottom panels are for the same simulations, but show the rescaled Kretschmann scalar $(\tau_*-\tau)I^{1/4}$ along null slices emitted from the center, against the similarity-adapted retarded time coordinate $T_\text{null}$, and the similarity-adapted affine parameter $\lambda$. We chose $\tau_*=5.4$ for both data sets.  To relate the top and bottom panels, the dotted line in the top right panel represents $T_{\rm null}=0$ in the bottom right panel, and the dashed line in the top right panel represents $\lambda=0.5$ in the bottom right panel.}
    \label{fig:kretsch_brill_both}
\end{figure*}

Throughout this paper we focus on axisymmetric Brill waves \cite{Bri59}, for which the extrinsic curvature of the initial data vanishes, and the initial spatial metric
\begin{equation}
 dl^2 = \psi^4 \left( e^q ( d\rho^2 + dz^2 ) + \rho^2 d \varphi^2 \right)
\end{equation}
is constructed from the seed function
\begin{equation} \label{Brill_seed}
    q = A\, \frac{r^2 \sin^2 \theta}{\sigma^2} e^{- (r / \sigma)^2}, 
\end{equation}
where $A$ is the amplitude (which can be positive or negative), $\rho = (x^2 + y^2)^{1/2} = r \sin \theta$ the distance from the symmetry axis, and $\sigma$ a constant with dimension of length. In the following we report all dimensional quantities in units of $\sigma$. Given $q$, we compute the conformal factor $\psi$ by solving
\begin{equation}
    \nabla^2 \psi = - \frac{\psi}{8} \left( \frac{\partial^2 q}{\partial \rho^2}  + \frac{\partial^2 q}{\partial z^2} \right),  
\end{equation}
where $\nabla^2$ is the flat-space Laplace operator.  

As a quantitative, gauge-independent comparison between our three codes we show in Fig.~\ref{fig:Brill_comparison} the maximum value $I_{\rm max}$ of the Kretschmann scalar (\ref{Kretschmann}) encountered in subcritical evolutions as a function of $A$.  For a CSS critical solution, one would expect $I_{\rm max} \simeq |A - A_*|^{-4\gamma}$ \cite{GarD98}, with the same $\gamma$ as for the black hole mass, compare (\ref{mass_scaling}), while, for a DSS critical solution, this simple power-law would have superimposed a ``wiggle" that is periodic in $\ln|A-A_*|$~\cite{Gun97,HodP97}.  For $A < 0$ data, which result in oblate geometries \cite{Hiletal13}, we adopt $A_* = -3.509144$ for the \prague\ code, $A_* = -3.509091$ for \bamps, and $A_* = -3.5088$ for \sphGR\ in Fig.~\ref{fig:Brill_comparison} and find very good agreement, demonstrating that all three codes predict very similar values for the critical amplitude $A_* \simeq -3.509$.   The resulting curves do display a wiggle about an approximate power-law, but without clear periodicity, suggesting that the threshold solution does not tend to exact DSS at least at our level of fine-tuning.

Brill waves with $A > 0$, which result in prolate geometries, pose more of a computational challenge \cite{Hiletal13,LedK21}. In Fig.~\ref{fig:Brill_comparison}, the choice $A_* = 4.696695$ for all three codes (consistent with \cite{HilWB17}) again results in very good agreement.  It is difficult to identify a power-law or a periodic wiggle, however, in particular since for $A_* - A \gtrsim 0.2$ the maxima $I_{\rm max}$ occur in the initial data, and hence provide no information about the threshold solution.  Our available data are nevertheless consistent with curvature scaling.

In order to analyze the threshold solution for the $A<0$ initial data directly, we show in the top row of Fig.~\ref{fig:kretsch_brill_both} values of the Kretschmann scalar $I$ on the symmetry axis for a near-critical evolution.  We include results from both the \bamps\ (left panel) and the \prague\ (right panel) codes, which allowed us the best fine-tuning to the threshold solution, displaying $I$ as a function of proper distance $z_p$ from the center, along slices of constant coordinate time which are labelled by proper time $\tau$ at the center.  Even though the two codes employ different slicing conditions, even these coordinate-dependent renderings show qualitative agreement.  In particular, we observe that regions with increasingly large curvature appear closer to the center and closer together in proper time at later times, suggesting a self-similarity with an accumulation point at the center.  

We next construct coordinates that allow us to explore any tentative self-similarity about the center in a fully gauge-invariant way.  Specifically, we first introduce the DSS-adapted retarded time coordinate 
\begin{equation} \label{self_sim_time}
    T_{\rm null} = - \ln ( \tau_* - \tau),
\end{equation}
where $\tau_*$ is the assumed proper time (at the center) of the accumulation event.  We then consider null geodesics emitted from the center, parameterized by an affine parameter $\lambda$.  We normalize $\lambda$ by choosing $\lambda = 0$ and $d \lambda / d \tau = (\tau_* - \tau)^{-1}$ at the center, so that, initially, $\lambda$ advances at the same rate as $T_{\rm null}$.  

The bottom row of Fig.~\ref{fig:kretsch_brill_both} shows the same data for the Kretschmann scalar $I$ as the top row, but as functions of $T_{\rm null}$ and $\lambda$.  The results from the two codes now agree very well.  Moreover, it is easy to identify patterns that repeat approximately periodically, even though this periodicity is not exact.  We conclude that the threshold solution for Brill waves with the negative $A$ seed-function (\ref{Brill_seed}) serves as an example of a vacuum threshold solution that is approximately DSS with an accumulation point at the center. Other examples include Teukolsky waves with seed functions used by AE and \cite{LedK21,BauGH23}.  From Fig.~\ref{fig:kretsch_brill_both} we crudely estimate the period to be $\Delta \simeq 0.6$, which is similar to the values reported by AE ($\Delta \simeq 0.5 - 0.6$, see their Table I) and \cite{BauGH23} ($\Delta \simeq 0.53$ for quadrupolar waves). 

\begin{figure}
    \centering
    \includegraphics[scale = 0.95]{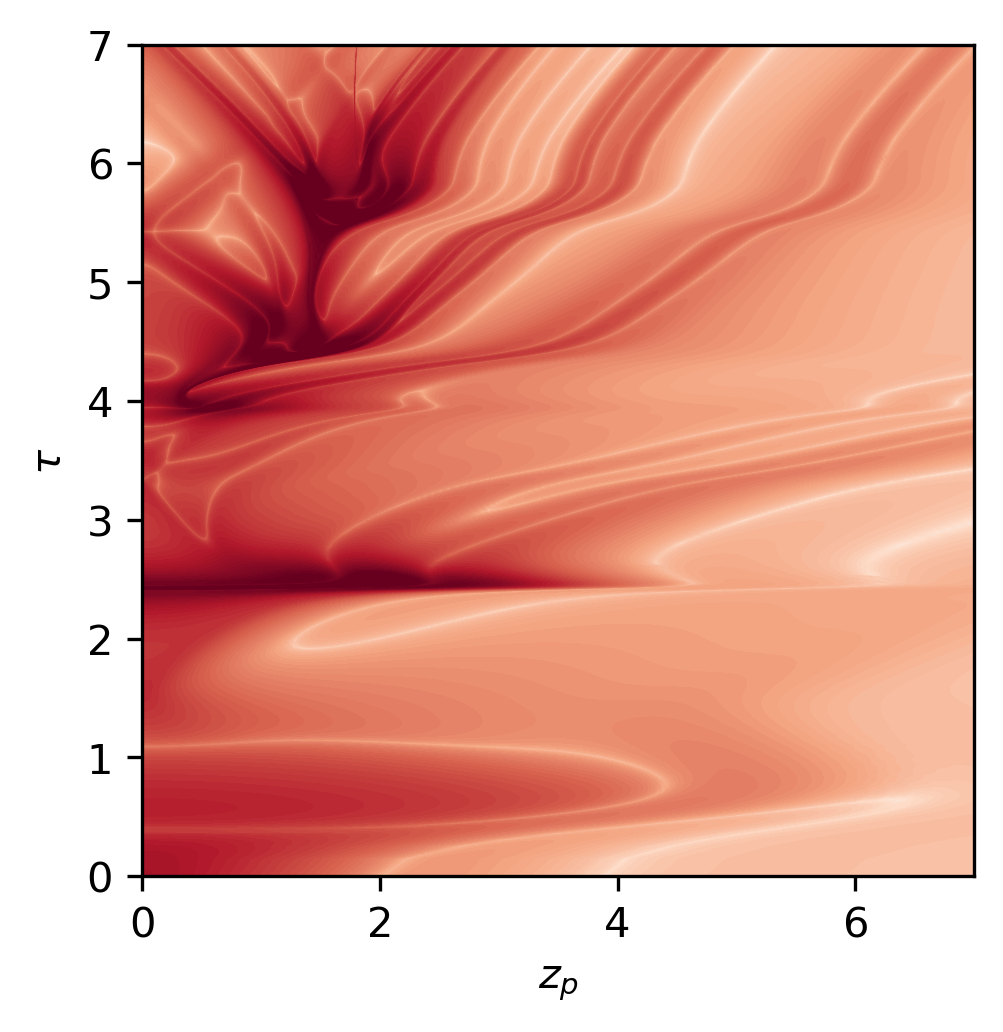}
        \caption{Similar to the top left panel in Fig.~\ref{fig:kretsch_brill_both}, but for a near-threshold $A > 0$ solution with $A=4.69667$.  We also plot $\log (I^{1/4})$ rather than $I^{1/4}$, and show \texttt{bamps} data only (but note that  coordinate-independent measures agree well with our other codes).  Unlike for the $A < 0$ data of Fig.~\ref{fig:kretsch_brill_both}, there is no evidence for a DSS with a single accumulation point at the center.}
    \label{fig:kretsch_brill_positive_a}
\end{figure}

As a demonstration that the threshold solution is not unique, however, we show in Fig.~\ref{fig:kretsch_brill_positive_a} the Kretschmann scalar for a near-critical evolution of Brill initial data with the seed function (\ref{Brill_seed}), but with $A > 0$.  Even from this Figure, which should be compared with the top-left panel of Fig.~\ref{fig:kretsch_brill_both}, it is clear that the solution does {\em not} feature an accumulation point at the center, and hence is distinct from the one for $A<0$.  It is possible that this solution features accumulation points away from the center but deciding this would require better fine-tuning than can presently be mustered.

To summarize, the qualitative and quantitative agreement among our three independent codes, together with the previous results presented in \cite{HilWB17,LedK21,SuaRCBH22,BauGH23}, allows us to draw several conclusions on the nature of critical collapse of vacuum gravitational waves.  Most importantly, there is no single, universal critical solution for the collapse of gravitational waves.  Rather, different families appear to lead to different threshold solutions with different scaling exponents and different locations of the accumulation point. For some families, the threshold solution appears to approach an approximately DSS spacetime with a single accumulation point.  The fact that our three codes agree quantitatively on the deviations from exact DSS, in particular the lack of clear periodicity, suggests that these are real, rather than numerical artifacts. We cannot rule out, of course, that these threshold solutions approach exact DSS with better fine-tuning. Conversely, in other families of initial data there is no evidence for DSS at the level of fine-tuning available to us.  It remains to be explored whether the complicated behavior we have begun to resolve numerically can be explained by multiple DSS solutions with one or more unstable modes, or whether vacuum collapse requires a more fundamental departure from our understanding of critical collapse in spherical symmetry.

%=========================================================
\acknowledgments
%=========================================================

TWB, CG, and DH gratefully acknowledge support through the Oberwolfach Research Fellows program at the Mathematisches Forschungsinstitut Oberwolfach in 2022.  Numerical simulations were performed at the Leibniz Supercomputing Centre (LRZ) [supported by projects~pn34vo and~pn36je], the Bowdoin Computational Grid, ``e-Infrastruktura CZ'' computer grid (project No. e-INFRA LM2018140), and Picasso from Málaga University funded by Red Española de Supercomputación (RES project FI-2023-1-0033). The work was supported in parts by the FCT (Portugal) IF Program~IF/00577/2015, PTDC/MAT-APL/30043/2017 and Project~No.~UIDB/00099/2020, Spanish Ministry of Science and Innovation (MCIN) and the Spanish Agencia Estatal de Investigación (AEI) grants PID2019-106416GB-I00/MCIN/AEI/10.13039/501100011033, RED2022-134204-E, RED2022-134411-T; the MCIN with funding from the European Union NextGenerationEU (PRTR-C17.I1); the FEDER Operational Program 2021-2027 of the Balearic Islands; the Comunitat Autònoma de les Illes Balears through the Direcció General de Política Universitaria i Recerca with funds from the Tourist Stay Tax Law ITS 2017-006 (PRD2018/23, PDR2020/11); the Conselleria de Fons Europeus, Universitat i Cultura del Govern de les Illes Balears; and EU COST Actions CA18108 and CA17137, by the Deutsche Forschungsgemeinschaft (DFG) under Grant No. 406116891 within the Research Training Group RTG 2522/1, by the Charles University Project GA UK No. 1176217, Czech Science Foundation Project No. GACR 21-11268S, and by National Science Foundation (NSF) grant PHY-2010394 to Bowdoin College.

% \bibliography{references}


\end{document}